\documentclass[aps,twocolumn,showpacs,superscriptaddress]{revtex4}
\usepackage{graphicx}\usepackage{epsfig}
\usepackage[psamsfonts]{amssymb}
\usepackage{amsfonts}\usepackage{amscd}
\usepackage{amsmath,amssymb}
\everymath{\displaystyle}
\usepackage{dcolumn}

\begin{document}
\thispagestyle{empty}
\date{\today}

\title{Asymptotics of the contour of the stationary phase
and efficient evaluation of the Mellin--Barnes integral for the
$F_3$ structure function}

\author{Aleksander V. Sidorov}
 \email{sidorov@theor.jinr.ru}
\affiliation{Bogoliubov Laboratory of Theoretical Physics,  \\ Joint
    Institute for Nuclear Research, Dubna, Moscow Region, 141980
    Russia}

 \author{Vasil I. Lashkevich}
 \affiliation{International Center for Advanced Studies, \\
Gomel State Technical University, Prospect Octiabria, 48, Gomel,  246746 Belarus}

 \author{Olga P. Solovtsova}
\email{olsol@theor.jinr.ru}
\affiliation{Bogoliubov Laboratory of Theoretical Physics,  \\ Joint
    Institute for Nuclear Research, Dubna, Moscow Region, 141980
    Russia}
 \affiliation{International Center for Advanced Studies, \\
Gomel State Technical University, Prospect Octiabria, 48, Gomel,  246746 Belarus}

\begin{abstract}
A new approximation is proposed for the contour of the stationary
phase of the Mellin--Barnes integrals in the case of its finite
asymptotic behavior as ${\rm Re} z\to -\infty $.
The efficiency of application of the proposed contour and
the quadratic approximation to the contour of the stationary phase
is compared by the example of the inverse Mellin transform for the structure function $F_3$.
It is shown that, although for a small number of terms $N$ in quadrature formulas
used to calculate integrals along these contours, the quadratic contour
is more efficient, but for $N>20$ the asymptotic stationary phase integration contour
gives better accuracy.
The case of the $Q^2$-dependence of the $F_3$ structure function is also considered.
\end{abstract}

\pacs{12.38.-t, 12.38.Bx, 13.60.Hb}

\maketitle

\section{Introduction}

The Mellin--Barnes (MB) integrals \cite{Bateman} are widely used at
high-energy physics. Recently, significant progress has been made in
the numerical computation of these integrals. The review of digital
packages is presented in the introduction to the paper
\cite{Gluza:2016fwh}. The list of physical tasks solved using the MB
integrals (two-loop massive Bhabha scattering in QED, in three-loop
massless form factors and static potentials, in massive two-loop QCD
form factors, in B-physics studies, in hadronic top-quark physics,
and for angular integrations in phase-space integrals) can be
supplemented by the problem of determining the structure functions
and parton distributions in x-space on the basis of their Mellin
moments.

The inverse Mellin transform method is widely used in calculations
related to deep inelastic scattering (DIS)
\cite{Gluck:1976iz,Gluck:1989ze,Graudenz:1995sk} for describing the
scaling violation in polarized and unpolarized structure functions
and fragmentation functions. The general expression for the inverse
Mellin transform is written as a contour integral in the complex $
z$-plane in the form
\begin{equation} \label{Mellin-1} f(x) =
\frac{1}{2\pi i} \int_{C} d{ z}\; x^{-{\rm z}} \tilde{f}({
z})\,,
\end{equation}
where the contour $C$ usually runs parallel to the imaginary axis,
to the right of the rightmost pole.
In case of the DIS, a function $\tilde{f}$ on the right-hand
side of expression (\ref{Mellin-1}) is
the moments of the structure function at some
fixed value of momentum transfer squared $Q^2_0$
and is usually expressed in terms
of the ratio of $\Gamma$-functions, like
the expression (\ref{Mellin-B})
for the Mellin moments of
the $F_3$-structure function,
which will be investigated in Sec.~IV.
Then, the integral (\ref{Mellin-1})
can be considered as a typical one-dimensional MB
integral.

The best efficiency in a numerical integration Eq.~(\ref{Mellin-1})
can be achieved on the contour of the stationary phase, where the
oscillations of the integrand are minimal. However, the solution of
the differential equation for the stationary phase contour and its
subsequent application to calculate the MB integral requires big
computing expenses (see, e.g., Ref.~\cite{Dubovyk:2017cqw}). Instead, it
is proposed to build such approximations of the stationary phase
contour that would allow the effective application of the quadrature
integration formulas \cite{Gluza:2016fwh}.
It can be said that the first attempt to construct an effective
approximation for the contour using the expansion of the integrand
at the saddle point was made by Kosower \cite{Kosower:1997hg} as applied to the calculation of parton distributions in the x-space.
We call this contour $C_K$.
Recently, it was proposed to construct the Pad$\acute{\rm e}$
approximation of the stationary phase contour, taking into account
the presence of a saddle point in the integrand and its asymptotic
behavior for large $z$ \cite{Gluza:2016fwh}. The effectiveness of
this approach is shown in the summary Table \cite{Gluza:2016fwh}.
However, according to this table, for the integrand $F_1(z,s)
=(-s)^{-z}\Gamma^3 (-z)\Gamma(1+z)/\Gamma(-2z)$ at $s = -1/20 $ the
relative accuracy $10^{-8}$ is achieved by taking into account 16
polynomials in the quadrature formula, whereas for the contour
$C_K$, we achieve the same accuracy with 12 terms. In this concrete
case,  the use of the integration contour $C_K$ is more effective than the use of the Pad$\acute{\rm e}$ approximation of the stationary phase contour constructed in Ref.~\cite{Gluza:2016fwh}.
When does the contour $C_K$, which does not take into account the
asymptotic behavior, allows us to obtain a reasonable result?
Will the advantages of this contour be preserved with increasing the required accuracy? The answer to these questions will prompt our consideration in this work.

Here, we propose a new approximation for the contour of
the stationary phase. The asymptotic behavior of the constructed
contour  coincides with the contour of the stationary phase as ${\rm
Re}z \to -\infty$. We consider a particular case of a finite
asymptotic behavior of the contour of the stationary phase in the
limit ${\rm Re}z \to -\infty$. The MB integral arising for the
$F_3$  structure function corresponds to this case.
First, the construction of the contour will be presented on a
simpler example of the MB integral whose value is known exactly. The
corresponding integral arising in Feynman diagrams was examined, for
example,  in Ref.~\cite{Gluza:2016fwh}, where it was denoted by
$I_5(s)$. Further, we consider in detail the application of the new
asymptotic contour in the calculation of the $F_3$ structure
function on the basis of their Mellin moments, and give a comparison with the results of applying the integration contour $C_K$.

It should be stressed that the choice of a contour of integration is
dictated by a physical task. For example, in the calculation of
massive diagrams it is usually enough to calculate the MB integral
once but with a high relative accuracy of $\sim10^{-10}-10^{-16}$.
In the case of finding the shape of the structure functions, the
accuracy of $ \sim 10^{-4}-10^{-5} $ is sufficient for the fit of
experimental data. However, in the process of fitting, the integral
has to be calculated many times (more than $10^6$ times), so the
computational speed that directly depends on the number of terms in
the quadrature formula $N$ is important.
It is known that the application of the contour proposed by Kosower is effective for a small number of $N$ terms in the quadrature formula when the nodes of the quadrature formula are located near the saddle point \cite {Kosower:1997hg,Sidorov:2017}. However, this contour considerably moves away from the contour of the stationary phase at large $|z|$.
One can expect that the advantages of the contour coinciding with
the asymptotic behavior of the exact contour of the stationary phase as ${\rm Re} z \to - \infty$ should be manifested for sufficiently
large values of $N$. We will try to find out at what value of
$N^{\star}$ occurs the ``change of the mode'', i.e. the use of the
contour coinciding with the asymptotic behavior of the contour of a
stationary phase becomes more efficient.

The paper is organized as follows. In Sec.~II, we start with
review of the basic expressions relating to the choice of the
integration contour in the MB integrals, according to the method
proposed in Ref.~\cite{Kosower:1997hg}.
In Sec.~III, we explain how a contour that coincides with the
asymptotic behavior of the stationary phase contour is constructed.
Here, we confine ourselves to integrals with finite asymptotic
behavior of the stationary phase contour as ${\rm Re} z\to -\infty
$. We present the example of the application of constructed contour
in the numerical calculation of well-known integral $I(s)$ whose
value is known exactly.
The case of the implementation of the MB integrals for evaluation of
the $F_3(x_{\tt B})$ structure function, where $x_{\tt B} $ is the
Bjorken variable, is given in Sec.~IV. We also investigate the
applicability of the asymptotic stationary phase integration contour
if the $Q^2$-dependence of the $F_3$ structure function is taken
into account.  Numerical estimates of the relative accuracy of the
reconstruction of the functions considered above using different
contours are given in Sec.~V. Summarizing comments are presented in
Sec.~VI.
In Appendix A, using the previously considered
integral $I(s)$, we perform an additional study for the
contour with finite asymptotic as~${\rm Re} z\to +\infty$.

\section{Construction of the $C_K$ integration contour}

Let us begin with the review of the main relations for creation of
the effective contour proposed by Kosower \cite{Kosower:1997hg}
based on the saddle-point method of the integrand in
Eq.~(\ref{Mellin-1}).
We use this contour, which recall that
is denoted by $C_K$, for comparison of the efficiency of the numerical evaluation of the integral (\ref{Mellin-1}) for different contours.

Following the saddle-point method one can rewrite expression
(\ref{Mellin-1}) as
\begin{equation} \label{Mellin-1a}
f(x) =\frac{1}{\pi} \int_{C{\;'}} {\rm Re} \left[ -i\, d {z}\;
F({z})\right] \,,
\end{equation}
where $F({z}) \equiv x^{-{z}} \tilde{f}({z})$,
$~{\it{C}{'}}$ means the contour running from the saddle point
$c_0$, where  $F'(c_0)=0$ (see Ref. \cite{Kosower:1997hg} for
additional details).
The complex variable ${z}$ is parameterized as
follows:
\begin{equation} \label{Mellin-z}
{z}(t)={x}(t)+i{y}(t) \quad \quad   ( {x}, \;{y}
~\text{real})
\end{equation}
with the conditions $~{x}(t_0=0)=c_0$ and ${y}(t_0)=0$.
Then, the expansion of $F(z(t))$ in a series around the saddle point
looks like
\begin{eqnarray}  \label{F-expansion}
~F(z(t)) \sim  F(c_0)
   &-& \frac{F''(c_0)}{2} \, t^2
 + \; \frac{1}{6} \biggl[ -i F^{(3)}(c_0)
 \\
  & + &   3i F''(c_0) x''(0)  \biggr]  t^3
  + \, \ldots \;~.
  \nonumber
\end{eqnarray}

The requirement that the imaginary part of the function
is equal to zero
\begin{equation} \label{ImF=0}
{\rm {Im}}F(z(t)) = 0 \;
\end{equation}
in any order of the $F$-expansion (\ref{F-expansion}) determines
the exact stationary-phase contour.

According to Eq.~(\ref{ImF=0}), to order ${\cal O}(t^3)$ the
contour is written as
\begin{equation} \label{Mellin-4}
{z}(t)= c_0+it+\frac{c_3}{2}\; t^2 \, ,
\end{equation}
where the coefficient $c_{3} = \frac{F^{(3)}(c_0)}{3  F''(c_0)}$.
Hence, to this order  we may write the integral
(\ref{Mellin-1a}) in the form:
\begin{equation} \label{Mellin-5}
f(x) =\frac{1}{\pi} \int\limits_{0}^{\infty} {\rm Re}
\left[ \left( 1- ic_{3} t \right)  F({z}(t))\right]
dt
\end{equation}
with the use of the quadratic integration contour $C_K$ given by Eq.~(\ref{Mellin-4}).

Next, putting $ ~t=c_2 {\sqrt{u}} \, $, one can introduce a new
variable  $u$, where $c_2 = \sqrt{2 F(c_0)/F''(c_0)}$, and the
integration over $u$ can be presented as
\begin{equation} \label{Mellin-6}
f(x) = \int\limits_{0}^{\infty} \frac{d u}{\sqrt{u}}\;
e^{-u} H_1(u)  \, ,
\end{equation}
where the function $H_1(u)$ up to order ${\cal O}(u)$ has the form
\begin{equation}
\label{h1(u)-1}
 H_1(u) = \frac{c_2}{2 \pi}\;{\rm Re} \left[
 e^u
 \left( 1-  ic_3c_2 \sqrt{u} \,
 \right)
 F
\left( {z}(u)  \right) \right]
\end{equation}
and the contour integration  $C_K$
in terms of
$u$ read as
\begin{equation}
{z}(u) = c_0 + i c_2\;\sqrt{u} + \frac{c_3}{2} \;c_2^2 \;u \,.
\label{Mellin-z-u}
\end{equation}

The function $H_1(u)$ up to order ${\cal O}(u^2)$ looks like
\begin{equation}
\label{h1(u)-2} {\overline{H}}_1(u) = \frac{c_2}{2 \pi}{\rm Re} \left[
e^u \left ( 1-  ic_3c_2 {\sqrt{u} - 4i c_6 c_2^3 u^{3/2}}\, \right) F
\left(\bar{z}\right) \right]
\end{equation}
and the integration contour is defined by the expression
\begin{equation}
\bar{z}(u)=c_0 + i c_2\;\sqrt{u} + \frac{c_3}{2} \;c_2^2
\;u+c_6 c_2^4 \; u^2
\label{Mellin-z-u2}
\end{equation}
with  the coefficients $c_{4} = \frac{c_3 F^{(4)}(c_0)}{12
F''(c_0)}$, $c_{5} = \frac{F^{(5)}(c_0)}{120 F''(c_0)}$, and $c_6
=\left(c_4-{c_5}-\frac{3}{8}c_3^3 \right)$. We designate this contour
by $\overline{C}_K$.
~It is obvious that if $c_6=0$, then expression (\ref{h1(u)-2}) turns
into Eq.~(\ref{h1(u)-1}).

Finally, one can see that the prefactor in front of the brackets in
Eq.~(\ref{Mellin-6}) corresponds exactly to the weight function for
the generalized Laguerre polynomials $L^{(-1/2)}_n(u)$.
Therefore,
one can expect that the application of the  generalized Gauss\---Laguerre quadrature formula
(see, e.~g., Ref.~\cite{Gauss-Laguerre})
\begin{equation} \label{Mellin-L}
 \int\limits_0^\infty {du\over\sqrt{u}}\;e^{-u}\; H_1(u) \simeq \sum_{j=1}^N w_j
H_1(u_j)
\end{equation}
can give a fast numerical evaluation of the integral
(\ref{Mellin-6}) in the lower orders of approximation.
This was the key achievement of the method
proposed by Kosower \cite{Kosower:1997hg}. Indeed, already the first
approximation $(N=1)$ gives relative accuracy of the parton
distributions reconstruction about a few percent (see, e.~g.,  Table in
Ref. \cite{Kosower:1997hg}). However, the number of points $N$
required for evaluating of the integral (\ref{Mellin-6}) using
Eq.~(\ref{Mellin-L}) with desirable accuracy depends on the integrand
function and remains the subject of a numerical study.

\section{Example for construction of asymptotic stationary phase contour}

In order to get a more detailed idea about the construction of the
asymptotic stationary phase integration contour, we begin with a
simple illustrative example of the MB integral, which can arise in
Feynman diagrams.  The asymptotics of the stationary phase contour
in this case tends to a finite limit, just as in the case which will
be considered below for the $F_3$ structure function.

Let us consider the integral~\footnote{Recall, in the paper
\cite{Gluza:2016fwh} the corresponding integral is denoted by
$I_5(s)$.}
\begin{equation}
{I}(s)  = \frac{1}{2\pi i}
\int\limits_{\delta-i\infty}^{\delta+i\infty}
dz\; (-s)^{-z}
\frac{\Gamma^3(-z)\Gamma(1+z)}{\Gamma(-2z)\Gamma(1-z)\Gamma(2+z)} \,
\label{M-B-I5}
\end{equation}
with the fundamental strip  $\delta\in(-1,0)$ in the region
$0<-s<4$. This integral can be evaluated analytically with result: $
I (s) = - s \, $.

We present the integrand as
\begin{eqnarray}
\Phi(z) \equiv e^{z\omega}
\frac{\Gamma^3(-z)\Gamma(1+z)}{\Gamma(-2z)\Gamma(1-z)\Gamma(2+z)} \,
, \label{exp_I}
\end{eqnarray}
where $\omega=-\ln(-s)$, and  using known asymptotic expressions $~
\Gamma(z) \sim  \sqrt{2\pi}\;  e^{-z} z^{z-{1}/{2}} \, , ~{z
\rightarrow\infty} \, ,$ and
\begin{equation}
\lim_{z \rightarrow\infty}{\frac{\Gamma(a+z)}{\Gamma(z)z^a}}=1 \,,
~|\arg{z}|<\pi-\epsilon\,,
 \label{assGammaz}
\end{equation}
we get the integrand  asymptotic behavior
\begin{eqnarray} \nonumber
\Phi(z) &\sim &|z|^{-5/2}\exp \left[ \left( \omega+ \ln 4 \right) z
+ i \; \frac{3}{2}\; \pi\;  {\rm sign}(y) \right.  \,
\\
 &  & \left.  \quad \quad  \quad \quad -i\frac{5}{2}\arg{z}\right] \, , ~~z=x+iy.
\label{assexpM3}
\end{eqnarray}
Whence from the zero phase condition ${\rm Im}\ln \left[\Phi(z)
\right]  = 0 \,$,  we obtain that $\arg{z} = {2}[\left(\omega+ \ln 4
\right)y+\frac{3}{2}\pi\; {\rm sign}(y)]/5 \, $ and the asymptotic
behavior of the zero-phase contour is determined by the equation
\begin{eqnarray}
x_{as}(y)=y {\rm ctg} \left[ {\frac{2}{5} \left( \left(\omega+ \ln 4
\right) y +\frac{3}{2}\pi\; {\rm sign}(y) \right)} \right] \,.
\label{ass_behavior}
\end{eqnarray}
From this equation it follows that in the asymptotic region as ${\rm
Re} z \to -\infty$ the zero-phase contour, which is a contour of
stationary phase  $C_{st}$, tends to lines parallel to the real axis
\begin{eqnarray}
y_{as}=\,\frac{~\pi}{\omega+ \ln 4} \; {\rm sign}(y) \,.
\label{ass_limits}
\end{eqnarray}
Thus, the asymptotic stationary phase contour $C_{as}$ in the
complex $z$-plane takes the form
\begin{eqnarray}
{z}_{as}(y) = x_{as}(y)+ i y \,.
 \label{contour_as}
\end{eqnarray}

Using this contour, we can represent the original integral (\ref{M-B-I5}) as
\begin{equation} \label{Mellin-I_5}
{I}(s) = \int\limits_0^{|y_{as}|} dy H_2(y) \,,
\end{equation}
where the function $H_2(y)$ is given by
\begin{equation}
\label{h(u)}
 H_2(y) =  {\rm Re} \left[ \left(1 -i
\frac{  d\; x_{as}(y)}{dy} \right) \Phi({z}_{as}(y)) \right] /
\pi  \, ,
\end{equation}
and, finally, we have
\begin{equation} \label{H2}
\int\limits_0^{|y_{as}|} dy H_2(y) \simeq
\frac{|y_{as}|}{2}\sum_{j=1}^N w_j H_2(y_j) \, ,
\end{equation}
where $y_j=\frac{|y_{as}|}{2}(x_j+1)$, $x_j$ are the roots of the
Legendre polynomials $P_n(x)$ with normalization $P_n(1)=1$, and
weight coefficients $w_j=\frac{2}{(1-x_j^2)[P^{'}_n (x_j)]^2 }$.

Note that in practical calculations of the integral
(\ref{Mellin-I_5}), it is advisable to shift
the asymptotic contour (\ref{contour_as})
parallel to the real axis to the saddle point $c_0$. Then, instead of
Eq.~(\ref{contour_as}), we get
\begin{eqnarray}
{z}_{as}(y) = x_{as}(y)+  \Delta c_0 + i y  \label{contour_as-shift}
\end{eqnarray}
where $ \Delta c_0 =c_0-c_{as}\,,~c_{as}\equiv x_{as}(y=0) \, $. For
the case under consideration, in accordance with
Eq.~(\ref{ass_behavior}), $x_{as}(y=0)=0$.

\begin{figure}[!h]
\begin{center}
{\includegraphics[width=7.0cm]{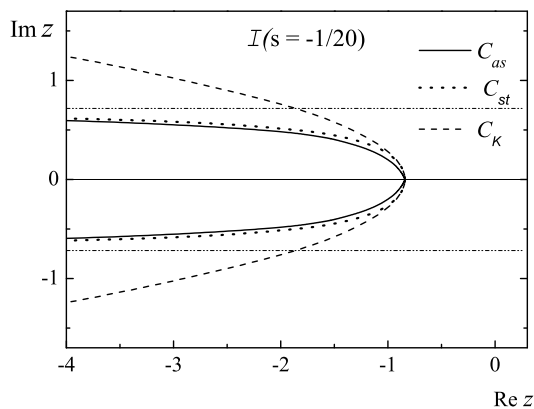}} \caption {The
efficient contours for the MB integral (\ref{M-B-I5}). The solid,
dotted, and dashed  curves indicate the contours $C_{as}$, $C_{st}$,
and $C_{K}$, respectively. Horizontal lines show the asymptotics
(\ref{ass_limits}). } \label{Fig-I5}
     \end{center}
\end{figure}

Comparing the shape of the contours, we also consider the exact
contour of the stationary phase, which can be found from the
differential equation
\begin{equation} \label{contour-s}
\frac{d {x}}{d {y}}=\frac{{\rm Re}\; \left\{
\partial_z  \ln \left[\Phi(z) \right]
\right\} ^\ast}{{\rm Im}\; \left\{
\partial_z  \ln \left[\Phi(z) \right]
\right\} ^\ast}
\end{equation}
with condition ${x}(0)=c_0 \, $. (The designation `$\ast$' means the
complex conjugation.) In the case under consideration, we used $x$
and $y $, since the solution $x = x (y)$ is uniquely determined. In
the general case, the equation for exact stationary-phase contour is
written in the parametric form using  Eq.~(\ref{Mellin-z}), see
Ref.~\cite{Gluza:2016fwh} for more details.

Figure~\ref{Fig-I5} shows the shape of the contours for the integral
$\;I(s)$ given by Eq.~(\ref{M-B-I5}) for a fixed value of $-s=1/20$.
The solid curve represents the asymptotic stationary phase contour
$C_{as}$ determined by Eq.~(\ref{contour_as-shift}), the dotted
curve corresponds to the exact contour of the stationary phase $C_{st}$ calculated by Eq.~(\ref{contour-s}), and the dashed curve is the contour $C_K$ corresponding to Eq.~(\ref{Mellin-4}).
The dash-dot-dotted horizontal lines denote the asymptotic limit of
the stationary phase contour determined by Eq.~(\ref{ass_limits}).
The main contribution to the integral $I(s)$ is given by a region
near the saddle point $c_0$, where the curves are very close to each
other. It can be found that the contours $C_{st}$ and $C_{as}$ are
close also in the asymptotic region ${\rm Re} z \to -\infty$, but
the contours $C_K$ and $\overline{C}_K$ quickly move away from the
contour $C_{st}$ (see Ref. \cite{Sidorov:2017} for details). Note
that for the considered value of $s$, the coefficient combination
$c_6c_2^4 $ in Eq.~(\ref{Mellin-z-u2}) is a small negative number
and the contour corresponding to Eq.~(\ref{Mellin-z-u2}) practically
coincides with the contour $C_{K}$. Therefore, we do not plot the
contour $\overline{C}_K$ in Fig.~\ref{Fig-I5}. The influence of the
coefficient $c_6$ for other region of $-s> 4$ is shown in the
Appendix~A.

\section{Asymptotic stationary phase integration contour for the $F_3$ structure function}

The reconstruction of the DIS structure function $F_3$ based on its
Mellin moments can be performed using the inverse Mellin
transformation (\ref{Mellin-1}), which reduces to the MB integral.
For an optimal calculation of this integral we construct the
efficient contours in exactly the same way as in the previous
section for the integral $I(s)$. It should be noted that the
accuracy of calculations of the structure function values can be
limited to 4--5 decimal places, which corresponds to the accuracy of
the experimental data.  At the same time, the method of calculating
of the $F_3$ structure function using the asymptotic stationary
phase contour makes it easy to obtain the $F_3$ values with high
accuracy.

\subsection{Contour of integration}

Let us consider the following parametrization  of the $F_3$
structure function~\footnote{In this subsection we omit the
$Q^2$-dependence of the $F_3$ structure function.}
 \begin{equation}  \label{xF3q0}
x_{\tt B}F_3(x_{\tt B}) = A\; x_{\tt B}^{\alpha}(1-x_{\tt B})^{\beta}\left(1+\gamma\;
x_{\tt B}\right)\,
\end{equation}
with values of the parameters found in Ref.~\cite{SS-F3-2}.

We can write the Mellin moments of the structure function
via the $ \Gamma$-functions:
\begin{eqnarray} \label{Mellin-B}
& & ~~~~~~~ M_3({z})=
\int\limits_{0}^{1}dx\;{x^{{z}-1}}xF_3(x) \\
& & = A \left[ \frac{\Gamma(\beta+1)\Gamma
(\alpha+{z})}{\Gamma(\alpha+\beta+1+{z})}
+\gamma\frac{\Gamma(\beta+1)\Gamma(\alpha+1+{z})}{\Gamma(\alpha+\beta+2+{z})}\right]~~~
\nonumber
\end{eqnarray}
and present the structure function in the form
\begin{equation} \label{Mellin-F3}
x_{\tt B}F_3(x_{\tt B}) = \frac{1}{2\pi i}\int_{C} d {z}\; x_{\tt B}^{- {z}}M_3({z})\, .
\end{equation}

Introducing the notation
\begin{equation}
\Phi(z)=e^{\omega_{\tt B} z}M_{3}(z)  \,, \label{expM3}
\end{equation}
where $\omega_{\tt B} \equiv -\ln(x_{\tt B})$, and using the
relation~(\ref{assGammaz}) one can get the asymptotic behavior
$\Phi(z)$ as $z\rightarrow\infty$
\begin{equation}
\Phi(z) \sim  e^{\omega_{\tt B} z} A \;\Gamma(\beta+1) \;
\frac{1+\gamma}{~{z}^{\beta+1}} \, .\label{ass_F3}
\end{equation}
Calculating the argument of the $\Phi$-function  and equating its
imaginary  part to zero, we arrive at the equation
\begin{equation}
\omega_{\tt B} y-(\beta+1)\;{\rm {arctg}}\;{\frac{y}{x}} = 0  \, .
\label{ass_equation}
\end{equation}
From this equation it follows that as $z$ tends to infinity, the
argument $z$ tends to $\pm\pi$ and we get the asymptotic behavior
\begin{equation}
x_{as}^{{\tt DIS}}(y)=y\; {\rm ctg} \left( \frac{\omega_{\tt B} \;
y}{\beta+1} \right) \,. \label{ass_behavior-DIS}
\end{equation}
Hence, in the asymptotic region  the contour
$C_{st}({\rm Re} z \to -\infty)$ tends to the finite limit
\begin{equation}
y_{as}^{{\tt DIS}}= \,\frac{(~\beta+1)\pi}{\omega_{\tt B}} \; {\rm
sign}(y) \,. \label{asDIS_limits}
\end{equation}
As a result, the asymptotics of the stationary phase contour are
parallel to the real axis. Note that the r.h.s. of
Eq.~(\ref{asDIS_limits}) depends on the Bjorken variable $x_{\tt B}$
and the parameter $\beta$ which relates to the shape of the
structure function at large $x_{\tt B}$-values. There is no
dependence on the parameters $\alpha$ and $\gamma$ contained in
Eq.~(\ref{xF3q0}). With the growth of $x_{\tt B}$ and $\beta$, the
width of the corridor between the asymptotic limits increases.
In accordance with expression (\ref{contour_as-shift}), we obtain that
the asymptotic stationary phase contour is determined by the expression
\begin{eqnarray}
z_{as}^{{\tt DIS}} = y\; {\rm ctg} \left( \frac{\omega_{\tt B} \;
y}{\beta+1} \right)  + c_0 - \frac{\beta+1}{\omega_{\tt B}} +i y
\, . \label{contour_DIS}
\end{eqnarray}

To calculate the $F_3$-structure function numerically, we can use
the expression (\ref{Mellin-I_5}), which is now read as
\begin{equation} \label{F3}
x_{\tt B}F_3(x_{\tt B}) = \int\limits_0^{|y_{as}^{{\tt DIS}}|}  dy
H_2^{{\tt DIS}}(y) \,,
\end{equation}
and also Eqs.~(\ref{h(u)}), (\ref{H2}), and (\ref{expM3}), replacing
$y_{as} \Rightarrow y_{as}^{{\tt DIS}}$ and $z_{as} \Rightarrow
z_{as}^{{\tt DIS}}$.

\begin{figure}[!ht]
\begin{center}
{\includegraphics[width=7.5cm]{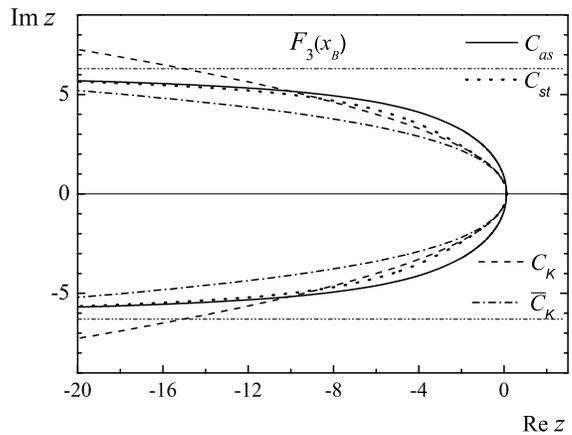}} \caption{The
comparison of the efficient contours for the $F_3$ structure
function at $x_{\tt B}=0.1$. The notations are the same as in
Fig.~\ref{Fig-I5}. The contour $\overline{C}_K$ is shown as the
dash-dotted curve. The horizontal lines  correspond to the
asymptotics (\ref{asDIS_limits}). }
 \label{Fig-F3}
     \end{center}
\end{figure}

Figure~\ref{Fig-F3} shows the shape of a set of contours  $C_{as}$
(solid curve), $C_{st}$ (dotted),  $C_K$ (dashed) and
$\overline{C}_K$ (dash-dotted) for a fixed value of $x_{\tt B}=0.1$.
In the calculation, we used the parameter values in
Eq.~(\ref{xF3q0}) that were found in Ref.~\cite{SS-F3-2} at fixed
$Q^2=Q_0^2=3$ GeV$^2$.

One can see that, qualitatively, the behavior of the curves in
Fig.~\ref{Fig-F3} is similar to the behavior in Fig.~\ref{Fig-I5}.
There is some difference between the shapes of the contours $C_{st}$
and $C_{as}$ in the vicinity of the saddle point.
However, in the limit ${\rm Re} z\to -\infty $ the contours $C_{st}$
and  $C_{as}$ coincide,  while  the contours $C_K$ and
$\overline{C}_K$ move away from the contour $C_{st}$ in this limit.

\subsection{$Q^2$ evolution and the contour of integration}

Let us discuss the change of the  behavior of the contour $C_{as}$
with a change of the momentum transfer squared $Q^2$.

The perturbative $Q^2$-evolution of the Mellin moments is well known
(see, e.g., Refs.~\cite{Buras:1979yt,Gluck:1989ze}), and in the
non-singlet case in the leading order ($LO$)  is given by the
formula
\begin{eqnarray} \label{evol-PT}
 & & M_3(z,Q^2) = M_3(z,Q^2_0) \exp \left\{ \Delta(Q^2) \right.\\
 & &  \left.  ~~~\times \left[\,
\frac{1}{2\,(z\,+\,1)\,(z\,+\,2)}\,+\,\frac{3}{4}\,-\,
\gamma-\psi(z+2)\,\right]  \right\} \,. \nonumber
\end{eqnarray}
Here
\begin{equation}
\label{DeltaQ2} \Delta(Q^2)\equiv \frac{16}{33\,-\,2n_f} \, \ln
\left[ \frac{~\alpha_{s}^{LO}(Q_0^{2})} {~\alpha_{s}^{LO}(Q^{2})}
\right] \, ,
\end{equation}
$\gamma$ is the Euler constant, $\psi$ is the usual logarithmic
derivative of the $ \Gamma$-function, $n_f$ is the number of active
flavors, and $\alpha_{s}^{LO}$ is the $LO$ running coupling.

Finding the asymptotic behavior of the integrand $\Phi(z,Q^2)\equiv
x_{\tt B}^{- {z}}M_3(z,Q^2)$ as $|z|\rightarrow\infty$, analogously
to how it was made before for a fixed of value $Q^2$, and also taking
into account an asymptotic behavior of the $\psi$ function, we obtain
\begin{equation}
\Phi(z,Q^{2}) \sim \exp{ \left\{ \omega_{\tt B} z - \left[ (\beta+1)
+ \Delta(Q^2) \right] \ln z \right\} } . \label{assexpDISQ2}
\end{equation}

\begin{figure}[!ht]
\begin{center}
{\includegraphics[width=6.5cm]{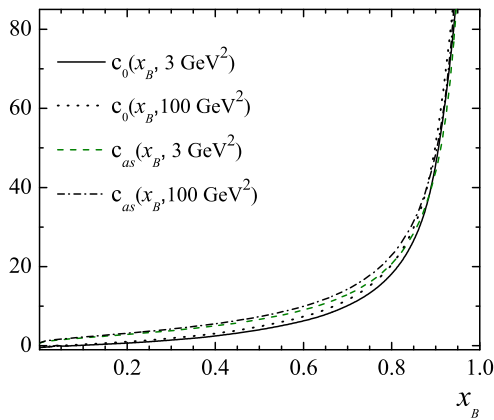}} \caption{
The comparison of the behavior of the saddle point  $c_0$
and the point $c_{as}$ versus of $x_{\tt B}$ at two fixed
values of $Q^2$.} \label{Fig-c0}
     \end{center}
\end{figure}

Next, from the zero phase condition follows that  $ \omega_{\tt B} y
-\left[(\beta+1) + \Delta(Q^2)\right]\arg{z} = 0 \,$. Therefore, the
asymptotic behavior of the contour as ${\rm Re}z \to -\infty$ is
determined by the formula
\begin{eqnarray}
x_{as}^{{\tt DIS}}(y,Q^2)=y {\rm ctg} \left[{\frac{\omega_{\tt B}
y}{(\beta+1)+\Delta(Q^2)}} \right] \, .\label{ass_behavior-Q2}
\end{eqnarray}
In the asymptotic region the contour $C_{as}$ tends to the finite
limiting value
\begin{eqnarray}
y_{as}^{{\tt DIS}}(Q^2)=\pm \, \frac{\left[
~(\beta+1)+\Delta(Q^2)\right] \pi}{\omega_{\tt B} }  \,.
\label{ass_limits-DIS}
\end{eqnarray}

Thus, the account of the $Q^2$-evolution of the Mellin moments of the
$F_3$ structure function comes down to replacement in the
expressions without evolution  $\beta \Rightarrow \beta
+\Delta(Q^2)$ and  using  Eq.~(\ref{F3})
$x_{as}^{{\tt DIS}}(y) \Rightarrow x_{as}^{{\tt DIS}}(y,Q^2)$ and $
y_ {as}^{{\tt DIS}} \Rightarrow y_{as}^{{\tt DIS}}(Q^2)$.

It is important to note that the expression defining the contour
$C_{as}$ in higher orders of the perturbation theory QCD will
coincides with the $LO$ expression with the replacement only of
$\alpha_{s}^{LO}(Q^2)$ in Eq.~(\ref{DeltaQ2}) by an expression for
the running coupling in the corresponding order of the perturbation
theory.

Figure~\ref{Fig-c0} shows the influence of the $Q^2$ evolution on a
values of the saddle point $c_0$ and the `asymptotic' point
$~c_{as}= x_{as}(y=0)$ in the case of the $F_3$ structure function.
In this figure $c_0$ and $c_{as}$ are shown as functions of the
Bjorken variable $x_{\tt B}$ at $Q^2=Q^2_0=3~{\rm GeV}^2$ (solid and
dashed lines for  $c_0$ and $c_{as}$, respectively) and
$Q^2=100~{\rm GeV}^2$ (dotted and dash-dotted lines).
One can see that there is no
significant difference in the behavior of the point $c_0$  with changing
of $Q^2$ as the solid and dotted curves are close to each other.
The same behavior is observed also for the point $c_{as}$.
It is interesting to note a sharp increase in the values of $c_0$
and $c_{as}$  if $x_{\tt B} > 0.5 $,
as well as the convergence of all curves at large values of~$x_{\tt B}$.

 \section{Numerical estimations of accuracy}

\begin{table}[t]
 \caption{The relative accuracy $\varepsilon(N)$ of numerical evaluation of
the MB integral (\ref{M-B-I5}) for different values of terms $N$ in
the sum (\ref{Mellin-L}), when the contour $C_K$ is used, and in the
sum (\ref{H2}) for the contour $C_{as}$.}

\begin{tabular}{c c c c c }
\hline \hline
\hspace{10mm} & \multicolumn{2}{ c }{~~~$-s=1/20$~~~} &
\multicolumn{2}{ c }{~~~$-s=2.0$~~~} \\ \hline
$N$ & $C_K$~~~ & $C_{as}$~~~ & ~~~$C_K$~~~ & $C_{as}$
\\  \hline
$16$ & $1.2 \times 10^{-8}~~$  & $1.3 \times 10^{-7}$    & $ ~~~ 1.2 \times 10^{-6}$ & $ ~6.5 \times 10^{-5}$ \\
$20$ & $6.7 \times 10^{-10}$ & $ ~ 8.5 \times 10^{-12}$  & $ ~~~ 4.0 \times 10^{-6}$ & $ ~3.9 \times 10^{-6}$ \\
$30$ & $1.2 \times 10^{-11}$ & $ ~ 5.7 \times 10^{-14}$  & $ ~~~ 5.1 \times 10^{-7}$ & $ ~1.2 \times 10^{-8}$ \\
$35$ & $5.8 \times 10^{-13}$ & $ ~ 5.7 \times 10^{-16}$  & $ ~~~ 1.4
\times 10^{-7}$ & $ ~5.8 \times 10^{-11}$
\\ \hline \hline
\end{tabular}
\label{tabl:1}
\end{table}


\begin{table*}
 \caption{The relative accuracy $\varepsilon(N)$ of
 the $x_{\tt  B}F_3(x_{\tt  B},Q_0^2)$ reconstruction
 for different values of terms $N$ in the sums (\ref{Mellin-L}) and (\ref{F3}),
 for the contours $C_K$ and $C_{as}$, respectively.}

\begin{ruledtabular}
\begin{tabular}{c c c c c c c }
\hspace{10mm} & \multicolumn{2}{ c }{$x_{\tt  B}=0.01$} &
\multicolumn{2}{ c }{$x_{\tt  B}=0.1$}& \multicolumn{2}{ c }{$x_{\tt  B}=0.5$}\\
\hline   $N$ & $C_K$ & $C_{as}$ & $C_K$ & $C_{as}$ & $C_K$ &
$C_{as}$   \\ \hline
$16$ & $2.5 \times 10^{-8}$  & $3.2 \times 10^{-7} $ & $1.3 \times 10^{-9} $ & $2.9 \times 10^{-8}$ & $1.0 \times 10^{-13}$  & $1.4 \times 10^{-11}$    \\
$20$ & $3.7 \times 10^{-10}$ & $4.1 \times 10^{-9} $ & $1.0 \times 10^{-10}$ & $1.6 \times 10^{-10}$ & $1.8 \times 10^{-17}$ & $4.6 \times 10^{-15}$  \\
$30$ & $2.7 \times 10^{-11}$ & $7.6 \times 10^{-13}$ & $5.7 \times 10^{-13}$ & $3.6 \times 10^{-16}$ & $3.8 \times 10^{-19}$ & $8.7 \times 10^{-21}$ \\
$35$ & $9.7 \times 10^{-13}$ & $1.5 \times 10^{-16}$ & $5.1 \times 10^{-14}$ & $1.5 \times 10^{-16}$ & $4.3 \times 10^{-20}$ & $3.6 \times 10^{-23}$\\
\end{tabular}
\end{ruledtabular}
\label{tabl:2}
\end{table*}

In this section, by numerically calculating the MB integrals given
by Eqs.~(\ref{M-B-I5}) and (\ref{Mellin-F3}), we investigate the
question at what values of polynomials $N$ in the corresponding quadrature formula
the advantages of the asymptotic contour $C_{as}$ are manifested,
in comparison with the quadratic approximation $C_K$ to the contour of the stationary phase.

The relative accuracy of a reconstruction is defined as usual
$\varepsilon(N) = |({f_N-f^{exact}})/{f^{exact}}|\, ,$ where $f_N$
is the sum given by Eq.~(\ref{Mellin-L}), when the contour $C_K$ is used, or  Eq.~(\ref{H2}) for the contour
$C_{as}$; $f^{exact}$ is the exact value of the corresponding integral.

The relative accuracy $\varepsilon(N)$ of calculating the $I(s)$
integral depending on number of terms $N$ in the sums
(\ref{Mellin-L}) and (\ref{H2}) is presented in Table~\ref{tabl:1}.
The result is given for $-s=1/20$ and $-s=2.0$.
The calculations for the $F_3$ structure function are presented in
Table~\ref{tabl:2} for the values $x_B=0.01,~0.1$, and $0.5$. In doing so, we use
the values of the parameters in  Eq.~(\ref{xF3q0}) that were
found in Ref.~\cite{SS-F3-2} at fixed $Q_0^2=3$ GeV$^2$.

Tables~\ref{tabl:1} and \ref{tabl:2} show similar results for the
relative accuracy $\varepsilon(N)$.
The contour $C_{K}$ works more efficiently for a small the number of terms
($N <20$) in the quadrature formulae (\ref{Mellin-L}) and (\ref{H2}),
as for a large $N>20$, the
result of the asymptotic contour $C_{as}$ is more accurate by 2--3
orders of magnitude. In addition, as can be seen from Table
\ref{tabl:2}, the high accuracy achieved at $N^{\ast}= 20$ exceeds
the experimental accuracy by several orders of magnitude. However,
as can be seen from Table~\ref{tabl:1}, the ``regime change``  for
the integral $I(-s=2.0)$ occurs at value of
$\varepsilon(N^{\ast})\simeq 4 \times 10^{-6}$. Such accuracy has
practical importance in the QFT calculations.

\begin{figure}[!h]
\begin{center}
{\includegraphics[width=8.0cm]{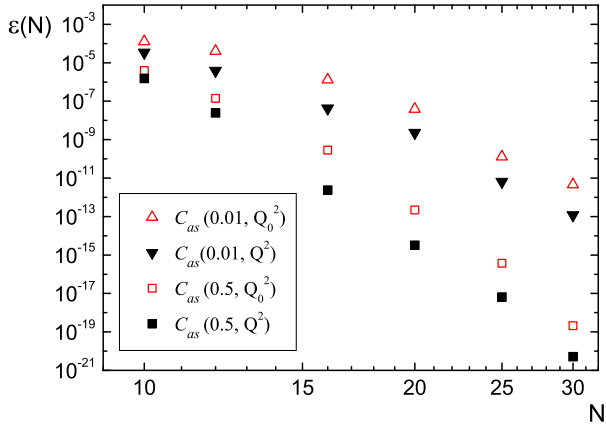}}\\
    \vspace*{0.1cm}
\caption{The comparison of the relative accuracy $\varepsilon(N)$ of
the $x_{\tt  B}F_3(x_{\tt  B},Q^2)$ reconstruction  versus the
number of terms $N$ in the sum (\ref{H2}) at
$x_{\tt B}=0.01$ (triangles) and $x_{\tt B}=0.5$ (squares)  using
the contour $C_{as}(x_{\tt B},Q_0^2)$ (open triangles and squares)
and $C_{as}(x_{\tt B},Q^2)$ (full triangles and squares) for
$Q_0^2=3$ GeV$^2$ and $Q^2=100$ GeV$^2$.} \label{Fig-3}
     \end{center}
\end{figure}

Figure \ref{Fig-3} shows the effect of the contour $Q^2$ evolution
according to Eq.~(\ref{ass_behavior-Q2}) for the $F_3$ structure function.
In this figure the relative accuracy $\varepsilon(N)$ is plotted as
a function of number  terms $N$ in the Gauss-Legendre quadrature
formula (\ref{H2}) at values of the Bjorken variable $x_{\tt B}=0.01$
(triangles) and $x_{\tt B}=0.8$ (squares) for the contours
$C_{as}(x_{\tt B},Q_0^2)$ (open triangles and squares) and
$C_{as}(x_{\tt B},Q^2)$ (full triangles and squares) for
$Q^2=Q_0^2=3$ GeV$^2$ and $Q^2=100$ GeV$^2$. For the running
coupling  $\alpha_{s}^{LO}(Q^2)$ we used a value $\Lambda_{{\rm
QCD}} = 363$~MeV and $n_f=4$ (see Ref.~\cite{SS-F3-2}). One can see
that the contour $C_{as}(x_{\tt B},Q^2)$ yields a more exact result;
however, this advantage is compensated when using the contour
$C_{as}(Q^2_0)$ if the number of terms $N$ is increased by 2--4
units.

\section{Conclusions}

We presented a new approximation for the construction of a contour
close to the contour of the stationary phase. A special case of the
finite  asymptotics of the stationary-phase contour in the limit
${\rm Re } z \to-\infty$ was considered as such an asymptotics
arises in the calculation of the MB integral that represents the
$F_3$ structure function in terms of its Mellin moments. The
proposed approximation reproduces the behavior of the contour zero
phase for large values of $|z|$ and has the form like
Eq.~(\ref{contour_DIS}) in the case of the DIS
structure functions.

It was compared the efficiency of application of
the asymptotic stationary phase contour $C_{as}$ and the contour
$C_{K}$, determined by the saddle-point method,
as described in Sec.~II, in calculating of the MB integrals (\ref{M-B-I5}) and (\ref{Mellin-F3}).
As expected, the contour $C_{K}$ turned out to be more effective for
a small number of $N$ terms in the Gauss\---Laguerre quadrature
formula, when the nodes of the quadrature formulas are located near
the saddle point \cite{Kosower:1997hg,Sidorov:2017}. The advantage
of the contour $C_{as}$ is manifested at large values of $N$. The
``regime change`` occurs at $N^{\ast} \approx 20$. For the region $
20<N \leqslant 35 $, the  contour $C_{as}$ gives a relative error of
2--3 orders of magnitude better than the contour $C_{K}$. With
increasing $N$, this gap increases too.

When the $ Q^2$-evolution of the structure function $ F_3 $ is taken
into account, the efficiency of the contours  $C_{as}{(Q_0^2)} $ and
$ C_{as}{(Q^2)} $ was compared.
Although the contour $C_{as}{(Q^2)}$ gives a more accurate result,
but this advantage is compensated by using the contour
$C_{as}{(Q_0^2)}$ if we increase the number of terms in the
quadrature formula only by 2--3 units. The contour $C_{as}{(Q_0^2)}$
can be considered as a universal one, that is, applicable for other
values of $Q^2$.

It should be emphasized that since the necessary accuracy is
achieved when using a small number of $N$ in the quadrature formulas
(\ref{Mellin-L}) or (\ref{Mellin-I_5}), the computer time for
calculating  the $F_3$ structure function by using
Eq.~(\ref{Mellin-F3}) with the efficient contours is significantly less than if one uses the
linear contours that are usually parallel to the imaginary axis, to
the right of the rightmost pole in the integrand, or a straight line
at an angle (see, e.g.,
Refs.~\cite{Gluck:1989ze,Graudenz:1995sk,deFlorian:2017lwf,Leader:2015hna}).

Here we have considered the $F_3$ structure function, which is the simplest among the DIS structure functions since it does not contain the contribution of gluons and sea quarks.
The parameterization of the shape of this structure function (\ref{xF3q0}) is typical and widely used in the QCD analysis of the structure functions. Our consideration can be  applied to the polarized nonsinglet
combination $\Delta q_3$, $\Delta q_8$ and the nonsinglet combination fragmentation function $D^{\pi^{+}}_{u_v}$.
The choice of the efficient contour for the singlet case can be performed along the same line, but requires more complicated formulas. This is the task for forthcoming investigation.

Our result can be useful in choosing an integration method in both
the one-dimensional case and the case of few-dimensional MB integrals if it is required to achieve relative accuracy unattainable in the integration by the Monte Carlo method.

\begin{acknowledgments}

This research was supported in part by the JINR-BelRFFR grant
F16D-004, JINR-Bulgaria Collaborative Grant, and by the RFBR Grant
16-02-00790.

The authors would like to thank Prof. J. Gluza for interest
in this work.

\end{acknowledgments}

\appendix
\section{O\lowercase{n numerical evaluation of} $I(s)$ \lowercase{for  $- s > 4$} }
\label{appA}

Let us turn to the MB integral (\ref{M-B-I5}) and a little discuss
the efficient contours for the case $-s > 4$. The exact result is
written as follows
\begin{eqnarray} \label{T_exct}
~~ I (s) = 2\ln (-s) &-& s \left( 1-\sqrt{1+\frac{4}{s}}
\right)  \\
 & + & 4 \ln \left[ \frac{1}{2} + \frac{1}{2} \sqrt{ 1+\frac{4}{s}}  \right] \,, ~- s > 4 \,. \nonumber
\end{eqnarray}

\begin{figure}[!ht]
\begin{center} \label{Fig-I5-2}
{\includegraphics[width=7.5cm]{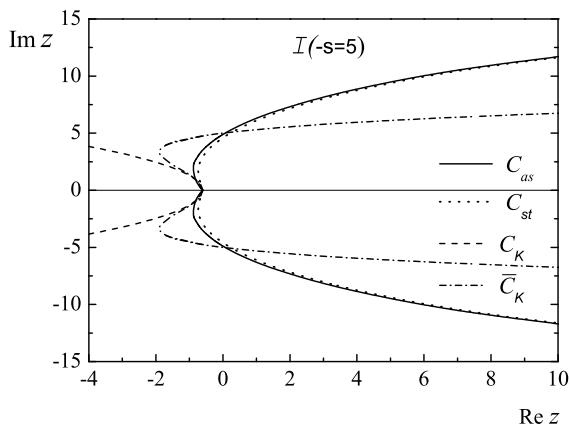}}
 \caption{The comparison of the efficient contours for the MB integral (\ref{M-B-I5})
in the region $-s > 4$. The notations are the same as in
Fig.~\ref{Fig-I5}. }
     \end{center}
\end{figure}

\noindent
This case is interesting for us,  because the exact contour of zero
phase $C_{st}$, see  Eq.~(\ref{contour-s}), go away from the
imaginary axis to the right.
The asymptotic behavior of the contour $C_{st}$ is defined by
Eq.~(\ref{ass_behavior}), as before, however, from this equation it
follows that in the asymptotic region as ${\rm Re} z \to +\infty$
the contour tends to another limit value
\begin{eqnarray}
y_{as}=\,\frac{3}{2}\frac{~\pi}{|\omega+ \ln 4|} \; {\rm sign}(y)
\,. \label{ass_limits-2}
\end{eqnarray}

Therefore, the numerical value of the integral $I(s)$ can be found
using the Gauss\---Legendre quadrature formula (\ref{H2}) in which
$y_ {as}$ is given by Eq.~(\ref{ass_limits-2}) and the contour
$C_{as}$ by Eqs.~(\ref{ass_behavior}) and(\ref{contour_as-shift}).

The contours $C_K$ and $\overline{C}_K$ are defined by the same way
as before, and the numerical value of the integral $I(s)$ is
determined using the Gauss\---Laguerre quadrature formula
(\ref{Mellin-L}).

Figure~\ref{Fig-I5-2} shows the shape of contours for fixed $-s=5$.
The solid curve corresponds to the contour $C_{as}$, the dotted
curve is the exact contour $C_{st}$, the dashed
curve is the contour $C_K$, and the dash-dotted is the contour
$\overline{C}_K$. From this figure it is clear that in calculating
the integral $I(s)$ using the contour ${C}_K$, difficulties
will arise, as this contour has an incorrect direction. Only the
first few terms in the sum (\ref{Mellin-L}) may give a reasonable
result, since the contour $C_{K}$ is close to the contour $C_{st}$
near the saddle point $c_0$.
Thus, for the case $-s > 4$ the contour $C_K$ it is impossible to
consider as efficient. At the same time, the contour
$\overline{C}_K$ works rather good, but applying it to achieve
relative accuracy, for example, $10^{-12}$, can be problematic.
The proposed contour $C_{as}$, whose a construction is quite simple,
reproduces the behavior of the exact contour $C_{st}$ well, and the
use of this contour can provide the required high relative accuracy.

\end{document}